\documentclass[pre,aps,twocolumn,10pt,showpacs,superscriptaddress]{revtex4-1}
\usepackage{upgreek} % to get upright mu in micron units (see below)
\usepackage{graphicx}
\usepackage{amsmath,amssymb}

\newcommand{\um}{\upmu\mathrm{m}}
\newcommand{\uN}{\upmu\mathrm{N}}
\newcommand{\half}{\frac{1}{2}}
\newcommand{\texthalf}{{\textstyle\half}}

\DeclareMathOperator{\sgn}{sgn}

\newcommand{\myav}[1]{\langle{#1}\rangle}

\newcommand{\Leibnizd}{{\mathrm d}}
\newcommand{\ds}{\Leibnizd s}
\newcommand{\dT}{\Leibnizd T}
\newcommand{\dy}{\Leibnizd y}
\newcommand{\dTds}{\frac{\dT}{\ds}}

\newcommand{\Tstar}{T^{*}}
\newcommand{\Tbreak}{T^{\ddag}} % or '+' or '\times' or something
\newcommand{\Tzero}{T^{(0)}}

\newcommand{\latin}[1]{{\itshape #1}}
\newcommand{\ie}{\latin{i.\,e.}}
\newcommand{\cf}{\latin{cf.}}
\newcommand{\etal}{\latin{et al.}}
\newcommand{\viceversa}{\latin{vice versa}}

\newcommand{\german}[1]{{\itshape #1}}
\newcommand{\ansatz}{\german{ansatz}}

\newcommand{\french}[1]{{\itshape #1}}
\newcommand{\ala}{\french{\`a la}}

\newcommand{\Eqref}[1]{Eq.\@ \eqref{#1}}
\newcommand{\Eqsref}[1]{Eqs.\@ \eqref{#1}}
\newcommand{\Figref}[1]{Fig.\@ \ref{#1}}

\begin{document}

\title{Why clothes don't fall apart: tension transmission in staple yarns}

\author{Patrick B. Warren}

\email{patrick.warren@unilever.com}

\affiliation{Unilever R\&D Port Sunlight, Quarry Road East, Bebington,
  Wirral, CH63 3JW, United Kingdom}

\author{Robin C. Ball}

\email{R.C.Ball@warwick.ac.uk}

\affiliation{Department of Physics, University of Warwick, Coventry,
  CV4 7AL, United Kingdom}

\author{Raymond E. Goldstein}

\email{R.E.Goldstein@damtp.cam.ac.uk}

\affiliation{Department of Applied Mathematics and Theoretical
  Physics, University of Cambridge, Wilberforce Road, Cambridge CB3
  0WA, United Kingdom}

\date{\today}

\begin{abstract}
The problem of how staple yarns transmit tension is addressed within
abstract models in which the Amontons-Coulomb friction laws yield a
linear programming (LP) problem for the tensions in the fiber
elements.  We find there is a percolation transition such that above
the percolation threshold the transmitted tension is in principle
unbounded, We determine that the mean slack in the LP constraints is a
suitable order parameter to characterize this supercritical state.  We
argue the mechanism is generic, and in practical terms corresponds to
a switch from a ductile to a brittle failure mode accompanied by a
significant increase in mechanical strength.
\end{abstract}

 % Tribology and mechanical contacts

\maketitle

In his celebrated \textit{Dialogues Concerning Two New Sciences},
Galileo identified a fascinating puzzle in the mechanics of ropes
\cite{galileo}.  His fictitious discussant Salviati asks: ``{\it How
  are fibers, each not more than two or three cubits in length, so
  tightly bound together in the case of a rope one hundred cubits long
  that great force is required to break it?}''  Galileo's answer to
this is to assert that ``the very act of twisting causes the threads to
bind one another in such a way that when the rope is stretched\,\dots
the fibers break rather than separate from each other.''  From a
modern perspective, we would say the mechanical integrity of ropes
derives from frictional contacts between fibers, and Galileo's rope
problem but one exemplar of a host of related frictional phenomena in fiber
assemblies, of which perhaps the canonical case is the `staple'
yarn \cite{HGB69, PHJY07, Pan92-93}.  Spun from fibers only
$2$--$3\,\mathrm{cm}$ long \cite{MH93}, such a yarn is nevertheless
patently capable of transmitting tension over indefinite distances.
Accompanying these seemingly innocuous puzzles is an even more
existential question: why don't clothes fall apart?  After all,
like Galileo's rope and the staple yarn, woven fabrics and sewn
garments are only held together by friction.

\begin{figure}[b]
  \centering
  \includegraphics[clip=true,width=0.9\columnwidth]{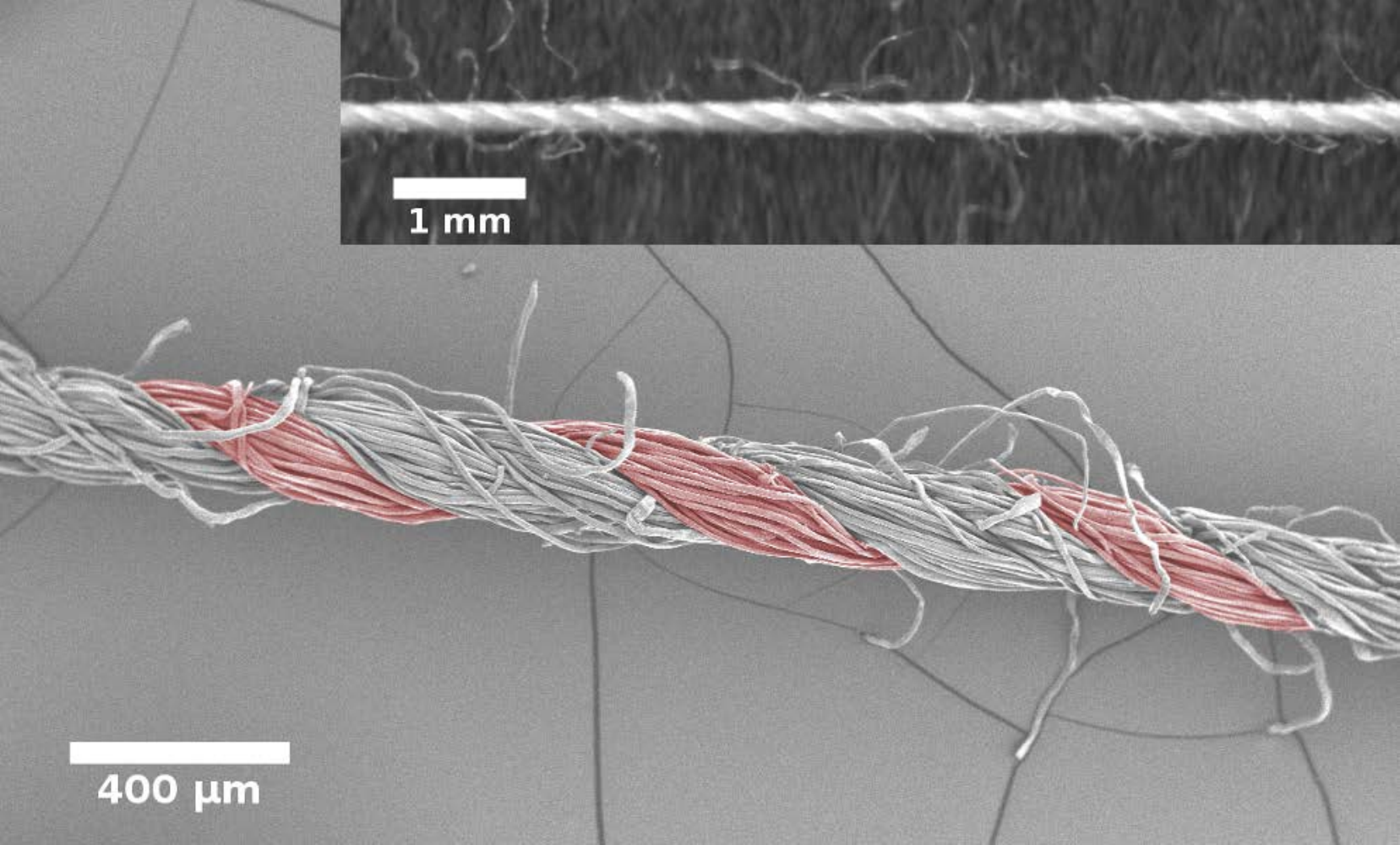}
  \caption{G\"utermann cotton sewing thread.  The composite 3-ply
    structure prevents untwisting under load.  One ply (yarn) has been
    artificially tinted to emphasize the structure.  Note the halo of
    stray fiber ends. Main image: SEM (Hitachi S-3400N); inset:
    flatbed scanner (Canon LiDE 220).\label{fig:image}}
\end{figure}

A typical yarn (\Figref{fig:image}) is $\sim 100$ fibers
in cross section, and there are likely several frictional contacts per
pitch length ($\sim\!100\,\um$), per fiber, hence we estimate
$\agt50$ contacts per fiber, and an overall contact density
$10^3$--$10^4\,\mathrm{cm}^{-1}$.  Clearly, the problem of tension
transmission in such a structure is a problem in statistical physics.
Here we introduce and explore a class of \emph{abstract yarn models}
which isolate the key frictional ingredients of such a problem.  Our
analysis supports the idea that given sufficient friction and contact
points, a random fiber assembly can in principle transmit an \emph{indefinitely
  large} tension, by means of a collective friction locking mechanism
that resembles a percolation transition.

The underlying premise is that normal forces acting between pairs of
fibers facilitate tension transfer between fibers.  The
Amontons-Coulomb friction laws \cite{GLG+04_short} then imply there is
an upper bound on the tension $\Delta T$ that can be transferred
before slip occurs.  Away from the fiber ends, the fibers are in a
tension-dominated regime even under modest loads \cite{tension}, hence
this tension transfer `cap' can be expressed as $|\Delta T|\le
\lambda\, T_m$ where $T_m$ is the mean tension in the notionally
over-wrapped fiber \cite{notion} and $\lambda$ is what we term a
tension transfer coefficient.  In the spirit of the approach we shall
take the transfer coefficients from a random distribution to reflect
the quenched disorder rather than attempting to solve the `inner'
elastic problem \cite{Stu61} for each pair of fibers. The key insight
is that if $\myav{\lambda}$ is large enough, this mechanism
`bootstraps' a percolation transition for tension transmission.

We accommodate the remnant yet singular effect of bending stiffness in
a lower bound to the tension in the fiber ends, which we estimate as
$\Tstar\alt1\,\mathrm{mN}$ \cite{tstar}.  This is illustrated in
\Figref{fig:yarn}a, where the tension in each section of fiber between
frictional contacts is shown building from zero at the free end.  What
$\Tstar$ means in practice is that we expect the percolation
transition to correspond to a switch from a `ductile' failure mode
where the yarn fails by fiber slippage, at around $\Tstar$ per fiber,
\cf\ \cite{sheets}, to a `brittle' failure mode where the failure 
mechanism is fiber breakage, at 
$\Tbreak\approx20$--$130\,\mathrm{mN}$ per fiber \cite{strength}.  As we shall
argue, twisting fibers together (\ala\ Galileo) pushes the assembly
over the percolation threshold, resulting in perhaps a hundred-fold
increase in the tensile strength.  Note that the scale separation
between $\Tstar$ and $\Tbreak$ means there is a significant
loading regime, of practical relevance, where tension can \emph{only}
be carried by the percolation mechanism identified in the present
work.
                  
Given the transfer coefficients, the problem of computing the set of
tensions $T_i$ in the fiber elements translates into a system of
linear inequalities which can be solved by techniques imported from
linear programming (LP).  From this perspective, the question of
whether the yarn transmits an arbitrarily large tension becomes a
\emph{linear satisfiability} problem.  In this form it is fairly easy
to show that $\Tstar$ is `irrelevant', in the language of
renormalisation group theory \cite{Fis98}, and as such we can carry
out all our calculations setting $\Tstar=0$ \cite{irrel}.  Our
approach shares elements with Bayman's `theory of hitches'
\cite{hitches}, although in our model a yarn is more akin to a random
continuous \emph{splice}, comprised of many short fibers, rather than
single-rope hitches.

\begin{figure}[t]
\centering
  \includegraphics[clip=true,width=0.95\linewidth]{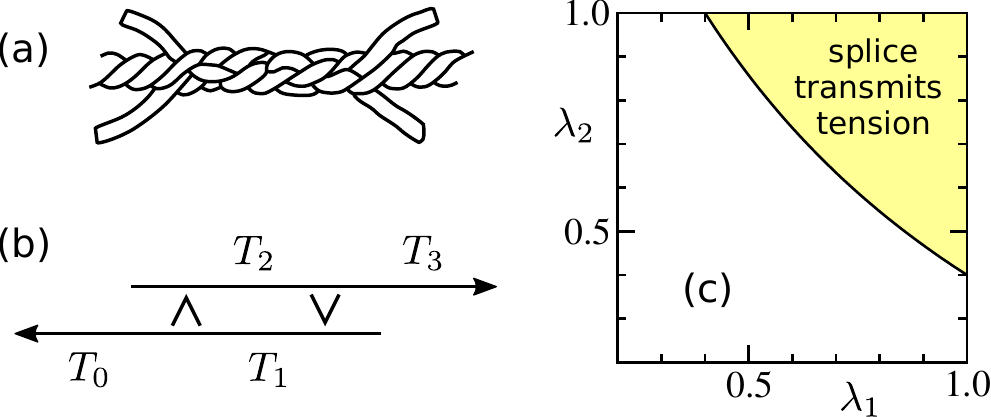}
  \caption{(a) A `short' splice between two laid ropes \cite{EB11}.
    (b) Schematic `toy' model of splice with labelled tensions (the
    $\vee$ and $\wedge$ shapes indicate the pinning direction).  (c)
    State space showing region (shaded) where tension can be
    transmitted. \label{fig:splice}}
\end{figure}

To explain the above we introduce a `toy' model of an actual splice,
shown in \Figref{fig:splice}.  Suppose that the tensions in the
various elements are as in \Figref{fig:splice}b, and the transfer
coefficients are $\lambda_1$ and $\lambda_2$.  Then,
%
% the '%' after the eqn label avoids a spurious space, see:
%https://tex.stackexchange.com/questions/54032/text-following-subequations-is-slightly-indented-if-a-label-is-used
%
\begin{subequations}
  \begin{align}
    &|T_1-T_0|\le\texthalf\lambda_1(T_0+T_1)\,,\quad T_0=T_1+T_2\,,\\[3pt]
    &|T_2-T_3|\le\texthalf\lambda_2(T_2+T_3)\,,\quad T_1+T_2=T_3\,,
  \end{align}
  \label{eq:1}%
\end{subequations}
where the inequalities are the tension transfer caps, and the
equalities are force balance constraints.  As mentioned, we simplify
by assuming tension-free fiber ends, and in this particular case make
a judicious choice for the over-wrapping direction (otherwise, the
splice would unravel).  We define the LP objective function
$z=\sum T_i$, and determine the percolation
threshold by requiring $z>0$.  This, together with
\Eqsref{eq:1} and the constraints $T_i\ge0$, specifies the LP  problem.

\begin{figure}[t]
  \centering
  \includegraphics[clip=true,width=0.95\linewidth]{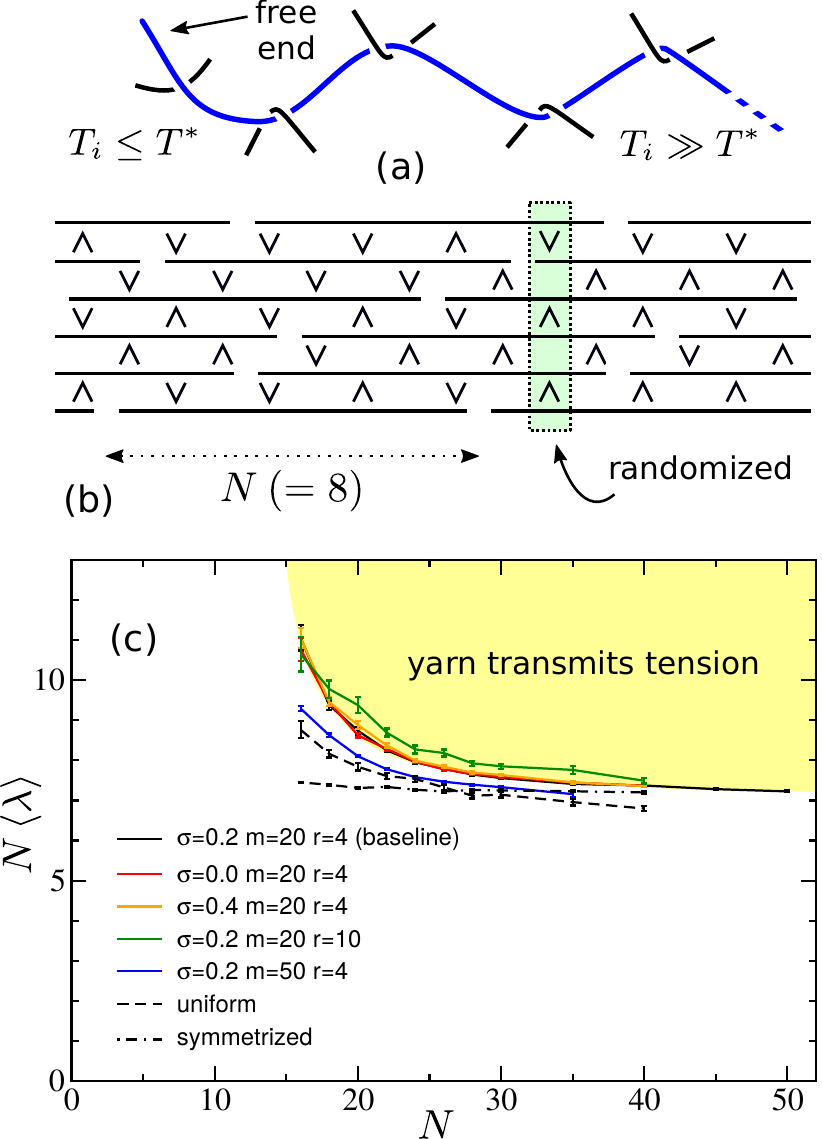}
  \caption{(a) A fiber meandering through the yarn structure
    accumulates tension by means of frictional contacts (angles are
    exaggerated in this schematic).  (b) Abstract yarn model in the
    notation of \Figref{fig:splice}b. For the baseline model the
    pinning assignments are randomly shuffled in each vertical column.
    (c) Critical value of $N\myav{\lambda}$ as a function of fiber
    length $N$, for different values of $m$, $r$, and the width
    $\sigma$ of the tension transfer coefficient distribution.  We
    also considered a `uniform' version without random vertical
    shuffling (but with random pinning directions); and a
    `symmetrized' version in which $T_m$ is the mean tension in
    \emph{all} fiber elements participating in a frictional
    contact. \label{fig:yarn}}
\end{figure}

This case can be solved by hand.  Defining $x=T_1/T_0$ and
$1-x=T_2/T_0$, with $0\le x\le1$, the caps yield
$x\ge(1-\half\lambda_1)/(1+\half\lambda_1)$ and
$x\le\lambda_2/(1+\half\lambda_2)$. A solution thus requires
$(1-\half\lambda_1) / (1+\half\lambda_1) \le \lambda_2 /
(1+\half\lambda_2)$, or $\lambda_1+\lambda_2 +
\frac{3}{2}\lambda_1\lambda_2 \ge 2$.  If this inequality is
satisfied, one is in a `locked' state where there are \emph{unbounded}
solutions with $z\to\infty$.  Intuitively (\Figref{fig:splice}c), such
solutions exist in the high-friction region.  Determining the value of
$x$ (\ie\ the individual tensions) in the supercritical locked state
is complex; it may depend on the history of loading, forces beyond
static friction, or the frictional contacts may adapt to the load, altering the
transfer coefficients.

\begin{figure}
\centering
  \includegraphics[clip=true,width=0.95\linewidth]{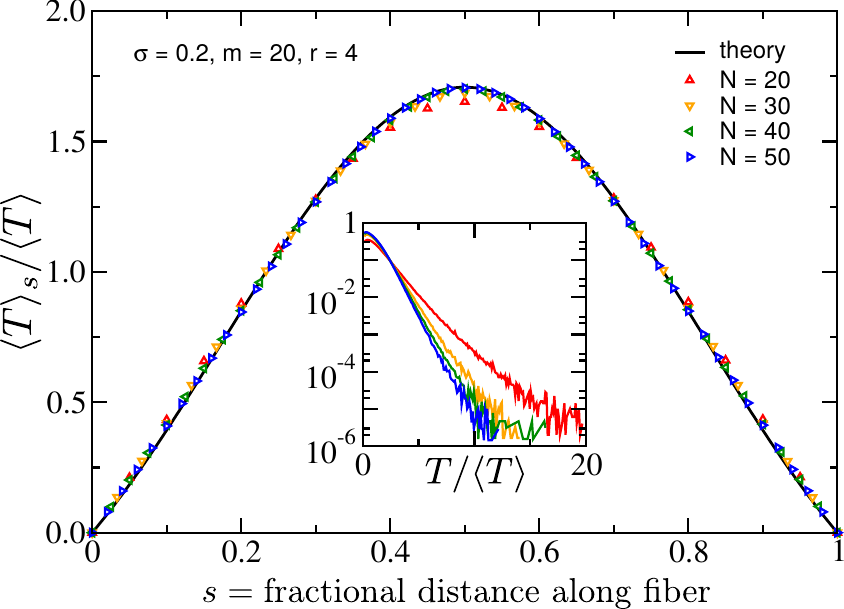}
  \caption{Mean tension as a function of distance along a fiber, for a
    baseline parameter set, computed just above the percolation
    transition.  Theoretical curve is the solution to \Eqref{eq:mft}
    at $\Lambda=\Lambda_c$.  The inset shows the distribution of
    tensions in individual fiber elements (\ie\ between pinning
    points).\label{fig:tensions}}
\end{figure}

We now describe the abstract yarn model which captures the essential
features of load transmission in fiber bundles by this mechanism.
Shown in \Figref{fig:yarn}a, it treats a yarn as a collection of
randomly overlapped near-parallel fibers, each of finite length $N$,
in units of the number of frictional contacts (pinning points).  The
structure comprises $m$ rows each of $r$ fibers, with a random
longitudinal offset in each row, and repeats periodically in the
transverse and longitudinal directions (in \Figref{fig:yarn}a, $m=6$,
$N=8$, and $r=2$).  The pinning assigments in each column shown in
\Figref{fig:yarn}a are randomly shuffled to mimic the random
meandering of fibers through the structure.  In the model there are
$(N+1)mr$ tensions $T_i\ge0$ and $\frac{3}{2}Nmr$ constraints coming
from $\half Nmr$ pinning points.  Thus for $N>2$ the problem is
potentially overconstrained, and solutions with $\sum T_i>0$ are
possible only if there is sufficient `slack' in the tension transfer
caps.  As mentioned, the tension transfer coefficients $\lambda_i$ are
independent and identically distributed random variables, with mean
$\myav{\lambda}$ and distribution relative width $\sigma$.  For each
structure we solve numerically for the onset of linear satisfiability
as $\myav{\lambda}$ increases, then average over $10^3$--$10^4$
samples.  \Figref{fig:yarn}b shows the dependence of the critical
$N\myav{\lambda}$ on the fiber length $N$.  In this representation the
results are insensitive to the model details, verifying our claim that
for sufficiently long fibers there is a generic percolation transition
in this model.
Solving for the $T_i$ just above the threshold yields
insight into the percolating system of forces. Thus
\Figref{fig:tensions} shows how tension in a fiber builds from zero at
the free ends, attaining a maximum in the middle, and the inset shows
the distribution of tensions in individual fiber elements.

For $N\agt30$ the percolation threshold is roughly constant at
$N\myav{\lambda}\approx7.3\pm0.2$.  In the tension-dominated regime,
and in the limit of a small turning angle $\theta$, the transfer
coefficient $\lambda\approx\mu\theta$ where $\mu$ is the fiber-fiber
friction coefficient.  If the critical $\myav{\lambda}\sim N^{-1}$,
this suggests that for $N\gg1$ one can interpret the percolation
threshold as a lower bound to the total fiber turning angle,
specifically $\mu\myav{\Theta}\agt 7$ where $\Theta\equiv N\theta$.
This quantifies Galileo's assertion about ropes, since
twisting fibers together builds $\Theta$, and parenthetically 
explains why \emph{spinning} is such an essential part of the
manufacturing process for yarns.  As a sanity check, for cotton
$\mu\approx0.3$--0.4 \cite{MH93, Pan92-93} and thus $\myav{\Theta}\agt
20$.  Table \ref{tab:thread} estimates the total fiber turning angle
for fibers in the yarns in \Figref{fig:image}, and it seems
this constraint is indeed comfortably met.

\begin{table}[t]
  \begin{ruledtabular}
    \begin{tabular}{llll}
      diameter & $d$ & 0.23--0.27  & mm\\
      apparent pitch & $\lambda$  & 0.30--0.39 & mm\\
      yarn curvature & $\kappa$ & 1.7--2.5 & mm$^{-1}$\\
      fiber length \cite{MH93} & $l$ & 20--30 & mm \\
      fiber turning angle & $\Theta$ & 30--70 & radians\\
    \end{tabular}
  \end{ruledtabular}
  \caption{Diameter and pitch (10--90 percentile ranges) of 3-ply
    G\"utermann sewing thread from imaging (\cf\ \Figref{fig:image}
    inset).  A proxy estimate for the fiber turning angle is
    $\Theta\approx\kappa l$. The yarn curvature $\kappa$ is estimated
    by modelling the ply centerline as a helix with diameter $d/2$ and
    pitch length $3\lambda$ \cite{helix}.\label{tab:thread}}
\end{table}

\begin{figure}[t]
\centering
  \includegraphics[clip=true,width=0.95\linewidth]{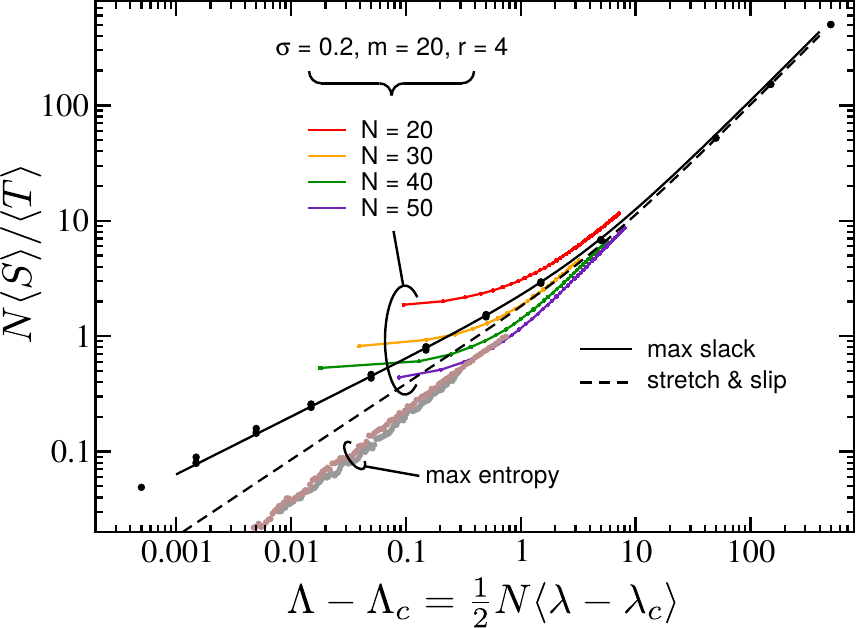}
  \caption{Normalised slack per fiber versus departure from
    criticality.  The `max slack' data points are numerical results
    from three samples each with one self-coupled fiber $N=2000$; the
    matching theoretical curve is \Eqref{eq:maxStheory} with no
    adjustable parameters.  The lower dashed line is for the more
    physical stretch-and-slip scenario.  The common asymptote at large
    $\Lambda$ is trivially $N\myav{S}=\Lambda\myav{T}$. The
    `max-entropy' data points show the slack averaged over the
    microcanonical ensemble of all admissible tensions at fixed
    $\myav{T}=1$, for two samples of a single self-coupled fiber with
    $N=2000$.  The lassoed rainbow data sets are from the yarn models
    of \Figref{fig:yarn}, averaged over edges of their solution cone:
    these are consistent with approaching the `max-entropy' results at
    large $N$.\label{fig:slack}}
\end{figure}

If we interpret tension transmission as a phase transition, it is
natural to seek an order parameter.  The load will not do, as the
problem as specified is homogeneous in the tensions.  Instead, for
each contact, we use the mean tension $T_m$ introduced previously to
define the \textit{slack}, $S=\lambda T_m-|\Delta T|$, as the amount by
which the tension transferred undershoots the friction limit
\cite{slack}.  The system-wide mean slack $\myav{S}$ is an order
parameter.  In the supercitical state there is generally not a unique
set of tensions (\cf\ selection of $x$ in the splice toy model),
rather there is a feasible solution \emph{space}, which in this case
is an open polytope: a convex, high-dimensional cone in the positive
hyper-quadrant of the space of fiber element tensions.  To compute
$\myav{S}$, we select a random edge of the solution cone, and average
over such edges.  The results (\Figref{fig:slack}) support the
notion of a second-order phase transition in the limit of long fibers
\cite{complex}, although there are significant finite-size effects.

To understand the nature of the percolation transition we now
develop a mean field theory for the tension $T(s)$ in a fiber
as a continuous function of fractional arc length $s$.  We
assume $N\gg1$ and correspondingly the tension transfer
coefficient $\lambda\ll1$.  The centerpiece of the
theory is the bi-dimensional function $\psi(s, s')$ which gives the
actual tension transfer between fibers in contact at $s$ and $s'$,
\ie\ as $\Delta T=\psi\lambda T_m$ with $|\psi|\le1$.  In these 
terms $T(s)$ satisfies the integro-differential equation 
\begin{equation}
  \dTds = \Lambda\int_0^1\! \ds'\,
  [ \psi(s,s')\,T(s')-\psi(s',s)\,T(s)]\,,
  \label{eq:mft}
\end{equation}
where $\Lambda=\half N\lambda$, noting that on
average there are $\half N$ frictional contacts of each type per
fiber. We additionally require $T(0)=T(1)=0$.  The mean slack is given
by $N\myav{S}=\Lambda\int\! \ds\int\! \ds'\, (1-|\psi|)\,T(s)$ where the
integral is over the square domain $(s,s')\in[0,1]\times[0,1]$.

Load bearing is enhanced by transferring tension to the fiber with
longer to go, so for $s<s'$ we transfer from $s'$ to $s$, and
\viceversa, and at criticality we must maximize this opportunity.
Thus as an \ansatz\ we choose $\psi(s, s')=\sgn(s'-s)$ (and
concomitantly, $\myav{S}=0$).  \Eqref{eq:mft} becomes
$\dT/\ds=\Lambda_c\int_0^1 \ds'\,\sgn(s'-s)\,[T(s)+T(s')]$.  The
resulting Sturm-Liouville-like problem can be solved, with normalized
solution $T(s) = 2x_0^2[1 - 2 x F(x)]$ where $x=x_0(2s-1)$,
$F(x)=\int_0^x\!\dy\, \exp{(y^2-x^2)}$ (Dawson's integral \cite{AS65}), and
$x_0\approx0.924$ solves $2x_0F(x_0)=1$.  The critical value
$\Lambda_c=4x_0^2\approx 3.416$.  For the tension profiles in
\Figref{fig:tensions}, precise agreement with the numerical results
is observed; we speculate the theory becomes exact in the limit of
long fibers.  The critical value yields $N\lambda_c\approx6.83$, in
good agreement with \Figref{fig:yarn}b, for long fibers.

Turning now to the supercritical behaviour, states with slack are
under-determined by the friction constraints alone.  The challenge is
to determine the fractional tension transfer $\psi(s,s')$ in
\Eqref{eq:mft} within the friction constraint that $|\psi|\le 1$. The
results are dependent on the choice of physics in the supercritical
state.  Here we sketch the main results \cite{longer}.  For example,
maximizing the slack selects $\psi(s,s')=0$ in a diagonal band
$|s'-s|<w$, whilst retaining the critical form $\psi(s,s')=\sgn(s'-s)$
outside this.  Treating the band as a perturbation ($w\ll 1$) recovers
the critical Sturm-Liouville-like problem with $\Lambda_c$ replaced by
${\Lambda}/{(1+\Lambda w^2)}$.  As a result the latter expression must
match $\Lambda_c$, leading to $1/\Lambda+w^2=1/\Lambda_c$.  For this
form of $\psi$ we readily find
\begin{equation}
  N\myav{S}=2w\Lambda \myav{T}
  =2\myav{T}\sqrt{(\Lambda/\Lambda_c)(\Lambda-\Lambda_c)}\,.
  \label{eq:maxStheory}
\end{equation}
To test this we used linear programming to solve the max-slack
condition for a single, long fiber transferring tension to itself.
The result (\Figref{fig:slack}) shows good agreement with the
theoretical curve in \Eqref{eq:maxStheory} over six decades
\cite{extrap}.

A more physical model is to presume that as we load the sample,
contact points displace affinely where they can within a `core' region
$|s-\half|< w$ and otherwise slide under locally critical conditions
(a `stretch-and-slip' model [\onlinecite{Pan92-93}b]).  This
means that all contacts associated with the `tails'
(\ie\ $|s-\half|\ge w$) are at their sliding condition with
$\psi(s,s')=\sgn(s'-s)$, including their contacts with points $s'$ in
the core. For points $s$ in the core we have affine deformation so
anticipate uniform strain leading to uniform tension.  This turns out
to be exactly compatible with \Eqref{eq:mft} choosing $\psi(s,s')=0$
when both $s$ and $s'$ are in the core, and noting that the
contributions to $\dT/\ds$ in the core from left and right tails
cancel each other out.  This leads to a second curve shown in
\Figref{fig:slack}.

Yet another possible scenario is to postulate that all supercritical
states within the allowed solution cone are equally likely (a
`max-entropy' model), akin to the microcanonical ensemble in
statistical mechanics, or the Edwards' conjecture in granular packings
\cite{EO89-MSR+17}.  A numerical investigation of this case is also
shown in \Figref{fig:slack}.

These possibilities lead to different values for the critical exponent
in $\myav{S}\sim(\Lambda-\Lambda_c)^\beta$, ranging from $\beta=1/2$
for the max-slack model, \Eqref{eq:maxStheory}, to $\beta\approx
0.75\pm0.05$ for the max-entropy case (fitting to a power law).  The
near-critical behavior of the
stretch-and-slip model is $w\propto (\Lambda-\Lambda_c)^{1/3}$ and 
$\myav{S}\propto w^2$, leading to the intermediate value $\beta={2}/{3}$.

To summarize, we propose a generic percolation transition as the
explanation for how staple yarns, woven fabrics, sewn garments (and
Galileo's rope) transmit tension over arbitrary distances.  Our
assertion is supported by the appearance of a transition in abstract
models, where the friction laws are recast as a linear satisfiability
problem.  This transition appears to be second-order, although the
critical exponents are dependent on physics beyond simple static
friction.  The abstract model may be generalised and applied in
various ways.  For example one can investigate fiber blends, with
applications to optimising the properties of functionalized sewing
threads.  In another direction, failure could be modeled by
iteratively breaking the most highly loaded fiber elements
(\Figref{fig:tensions} inset), \cf\ elastic fiber bundle models
\cite{PHC10}.  More generally, the LP approach to Amontons-Coulomb
friction problems may have applications in stress transmission in
granular media such as sand piles and grain silos.

\begin{acknowledgments}
We thank Simon Johnson for a critical reading of the manuscript,
Andrea Ferrante for helpful comments, and Jane Munro-Brown for the SEM
image used in \Figref{fig:image}.  We acknowledge an anonymous referee
for insightful questions regarding what happens near the fiber ends.
This work was supported in part by an Established Career Fellowship
from the EPSRC (REG).
\end{acknowledgments}

\vspace{12pt}\noindent{\small This arXiv preprint is the final author
  version (before copy editing) of ``Why Clothes Don't Fall Apart:
  Tension Transmission in Staple Yarns'', P. B. Warren, R. C. Ball and
  R. E. Goldstein, Phys.\ Rev.\ Lett.\ {\bf120}, 158001 (2018).}

\appendix

\section*{Appendix: Supplemental Material}
We provide a short proof that the inhomogeneous LP problem has
unbounded solutions if and only if the homogeneous LP problem has
unbounded solutions, thus the percolation threshold can be computed
with $\Tstar=0$.

Suppose that the \emph{homogeneous} LP problem is specified by
$T_i\ge0$, subject to tension transfer limits at mechanical contacts
$|\Delta T|\le\lambda T_m$, and $T_i=0$ at the fiber ends, with the
objective function $z=\sum T_i$.  The \emph{inhomogeneous} LP problem
is the same, but with $T_i\le\Tstar$ at the fiber ends.

First note that every solution to the homogeneous LP problem is also a
solution to the inhomogeneous LP problem, since setting $T_i=0$ in the
fibre ends satisfies $T_i\le\Tstar$.  Therefore if the homogeneous LP
problem has unbounded solutions, so does the inhomogeneous problem.
To prove the converse, let an unbounded class of solutions of the
inhomogeneous LP problem be
\begin{equation}
  T_i=\Tzero_i+\alpha R_i
\end{equation}
where $\alpha>0$.  The `ray' $R_i$ represents a direction in
$T_i$-space in which unbounded solutions exist.  Certainly, some of
the $R_i$ will vanish, but not all of them, since we know that if $z$
is unbounded at least one of the $T_i$ should be unbounded.  Therefore
solutions of this kind exists.  Also, if the lower bound to $\alpha$
is not zero, it can be absorbed into the definition of $\Tzero_i$.
Finally, we must have $R_i\ge0$ since if any $R_i$ was negative we
could violate the constraint $T_i\ge0$ by making $\alpha$ large
enough.

By substitution, and dividing through by $\alpha>0$, we find the $R_i$
satisfy
\begin{equation}
  \begin{split}
    &|\Delta R+\alpha^{-1}\Delta\Tzero|\le
    \lambda(R_m+\alpha^{-1}\Tzero_m)\,,\\[3pt]
    &R_i+\alpha^{-1}\Tzero_i\le
    \alpha^{-1}\Tstar\quad\text{(fibre ends)}\,,
  \end{split}
\end{equation}
where the notation is hopefully obvious.
Letting $\alpha\to\infty$ in this gives
\begin{equation}
  |\Delta R|\le\lambda R_m\,,\qquad
  R_i=0\quad\text{(fibre ends)}\,.\label{eq:S3}
\end{equation}
Thus the ray direction $R_i$ solves the homogeneous LP problem.  This
means that given unbounded class of solutions to the inhomogeneous LP
problem, one can construct a solution to the homogeneous LP problem. Then, 
since any positive multiple $\beta R_i$ ($\beta>0$) also solves
\Eqsref{eq:S3}, one has in fact constructed a class of unbounded
solutions to the homogeneous problem.  This proves the equivalence.


\begin{thebibliography}{24}
\bibitem [{ten()}]{tension}%
  \BibitemOpen
  \href@noop {} {}\bibinfo {note} {Tension-dominated means $T_i\gg\Tstar$ where
  $\Tstar$ is defined below; in this regime (yarn load $\agt$ few g\,wgt),
  bending stiffness can be safely ignored away from the fiber
  ends.}\BibitemShut {Stop}%
\bibitem [{not()}]{notion}%
  \BibitemOpen
  \href@noop {} {}\bibinfo {note} {The results are insensitive to the choice of
  which fiber is notionally overwrapped since this subtlety is erased by
  randomly shuffling the pinning directions. Also, we expect adjacent
  frictional contacts are elastically decoupled in the tension-dominated
  regime, justifying the choice to use independent and identically distributed
  tension transfer coefficients.}\BibitemShut {Stop}%
\bibitem [{tst()}]{tstar}%
  \BibitemOpen
  \href@noop {} {}\bibinfo {note} {The estimate derives from $\Tstar=\mu N$
  where the friction coefficient $\mu=O(1)$ for these purposes. The beam
  equation then gives the normal force $N\approx B/L^2$, where the bending
  stiffness $B\approx3$--$35\,\uN\,\mathrm{mm}^2$ and the fiber `crimp' length
  scale (or the distance between frictional contacts) we estimate as $L\agt
  100\,\um$. The bending stiffness in turn derives from the reported specific
  flexural rigidity of cotton,
  $0.53\,\mathrm{mN}\,\mathrm{mm}^2\!/\mathrm{tex}^2$ \cite{MH93}, in
  combination with the reported range of cotton fiber linear densities
  0.7--$2.3\,\mathrm{denier}$, where $1\,\mathrm{denier} =
  1\,\mathrm{g}/9000\,\mathrm{m} = 0.11\,\mathrm{tex}$.}\BibitemShut {Stop}%
\bibitem [{hel()}]{helix}%
  \BibitemOpen
  \href@noop {} {}\bibinfo {note} {That is to say, $\kappa = (d/4) /
  [(d/4)^2+(3\lambda/2\pi)^2]$.}\BibitemShut {Stop}%
\bibitem [{sla()}]{slack}%
  \BibitemOpen
  \href@noop {} {}\bibinfo {note} {This definition of slack is closely related
  to, but not exactly the same as, the LP notion of slack.}\BibitemShut {Stop}%
\bibitem [{com()}]{complex}%
  \BibitemOpen
  \href@noop {} {}\bibinfo {note} {For another slant on the problem we also
  monitored the number of interations of the LP simplex solver required to
  decide satisfiability as a measure of computational complexity. We find this
  exhibits a peak at the critical point, reminiscent of the behavior of the
  heat capacity at a conventional second-order phase transition.}\BibitemShut
  {Stop}%
\bibitem [{ext()}]{extrap}%
  \BibitemOpen
  \href@noop {} {}\bibinfo {note} {The success in extrapolating the
  perturbation calculation is explained by the fact that \Eqref{eq:maxStheory}
  asymptotes to $N\myav{S}/\Lambda\myav{T}=2\Lambda_c^{-1/2}=1/x_0\approx
  1.082$; this is \emph{almost} the correct large $\Lambda$ behaviour,
  $N\myav{S}/\Lambda\myav{T}=1$.}\BibitemShut {Stop}%
\end{thebibliography}
\end{document}